\documentclass[11pt]{article}
\usepackage{amsfonts}
\usepackage{amssymb}
\usepackage{amsmath}
\usepackage{amsthm}
\usepackage{amsfonts}
\usepackage{multicol}
\usepackage{amscd}
\usepackage{textcomp}
\usepackage{multirow}
\usepackage[english, activeacute]{babel}

\def\be{\begin{equation}}
\def\ee{\end{equation}}
\def\bea{\begin{eqnarray}}
\def\eea{\end{eqnarray}}

\newcommand{\sect}[1]{\setcounter{equation}{0}\section{#1}}
\newcommand{\subsect}[1]{\subsection{#1}}

\def\tfrac#1#2{ {\scriptstyle { \frac {#1}{#2}}}}

\def\1{\'{\i}}

\def\>#1{{\mathbf#1}}

\begin{document}

\thispagestyle{empty}


\ 
\vspace{0.5cm}

\begin{center}

{\Large{\sc{Integrable deformations of Lotka-Volterra systems}} }

\end{center}

\medskip

\begin{center} \'Angel Ballesteros, Alfonso Blasco and Fabio Musso
\end{center}

\begin{center} {\it {Departamento de F\1sica,  Universidad de Burgos, 
09001 Burgos, Spain}}

e-mail: angelb@ubu.es, ablasco@ubu.es, fmusso@ubu.es
\end{center}

  \medskip

\begin{abstract} 
\noindent
The Hamiltonian structure of a class of three-dimensional (3D) Lotka-Volterra (LV) equations is revisited from a novel point of view by showing that the quadratic Poisson structure underlying its integrability structure is just a real three-dimensional Poisson-Lie group.
As a consequence, the Poisson coalgebra map $\Delta^{(2)}$ that is given by the group multiplication provides the keystone for the explicit construction of a new family of 3N-dimensional integrable systems that, under certain constraints, contain N sets of deformed versions of the 3D LV equations.
Moreover, by considering the most generic Poisson-Lie structure on this group, a new two-parametric integrable perturbation of the 3D LV system through polynomial and rational perturbation terms is explicitly found.
 \end{abstract}

\bigskip\bigskip\bigskip\bigskip



\vfill
\newpage


\sect{Introduction}

The Lotka-Volterra (LV) system of differential equations 
\be
\dot{x}_i= x_i\left[\sum_{j=1}^{L}{a_{ij}\,x_j}+ d_i\right]
\qquad i=1,\dots,L
\label{LVN}
\ee
where $a_{ij}$ and $d_i$ are constants, were initially introduced for $L=2$ in order to model prey-predator and chemical reaction dynamics~\cite{Lotka,Volterra}, but nowadays it is well known that the generic system (\ref{LVN}) plays an outstanding role in many different nonlinear models with a wide range of applications (see, for instance~\cite{Plank, BF} and references therein).

In particular, the three-dimensional ($L=3$) LV system has attracted much attention, in many cases focused in the search for the specific values of the parameters $a_{ij}$ and $d_i$ for which a constant of the motion does exist (see~\cite{Bountis,CFG, Moulin, CF, CLl, CDamianou, ChrisDamianou}). Obviously, such integrability is guaranteed in case that the LV system could be obtained as the Hamilton equations $\dot{x}=\{x, \mathcal{H}\}$ provided by a Hamiltonian $\mathcal{H}$ and a  suitable three-dimensional (generalized) Poisson structure~\cite{Nutku, Damianou, GNutku, Plank2, Kerner, BF2, Puyun}. In particular, in~\cite{Nutku} it was shown that an integrable LV system previously described in~\cite{Moulin} and given by
\begin{eqnarray}
\dot{x}&=& x (C y +z +\lambda)\notag\\
\dot{y}&=& y (x+A z+ \mu)\label{abc}\\
\dot{z}&=& z(B x+y+\nu)
\notag
\end{eqnarray}
where
\be
A\,B\, C=-1
\qquad
\mbox{and}
\qquad
\nu=\mu B-\lambda A B
\ee
was bi-Hamiltonian. One of the Hamiltonian functions leading to (\ref{abc}) is given by
\be
\mathcal{H}=A\,B\,x+ y- A\,z+\nu\,\ln y-\mu\ln z
\label{hamab}
\ee
together with the  quadratic Poisson structure
  \be
\begin{array}{lll}
\{x,y\}=C\, xy\qquad & \{x,z\}= BC\,xz \qquad & \{y,z\}=- yz
\label{pnut}
\end{array}
\ee
which has the following Casimir function (that is indeed a constant of the motion for the LV system):
\be
\mathcal{C}=AB\ln x - B\ln y+\ln z.
\ee
Moreover, the second Hamiltonian structure for the same system is obtained by taking $\mathcal{C}$ as the Hamiltonian and by considering a different cubic Poisson bracket, for which $\mathcal{H}$ is just the Casimir function (see~\cite{Nutku} for details).

The aim of this paper is to generalize the abovementioned quadratic Hamiltonian structure of the LV equations by showing that it can be interpreted as a Poisson-Lie bracket on a certain Lie group. Therefore,  these 3D LV equations can be understood as an instance of integrable Poisson-Lie dynamics, and this property will enable us to get new integrable deformations of the LV equations in two different ways. Firstly, by using the group multiplication (which defines by construction a Poisson map $\Delta^{(2)}$ between two copies of the group manifold) in order to get an embedding of certain integrable perturbations of LV equations into a higher dimensional integrable system. Secondly, by looking for deformations of the underlying quadratic Poisson-Lie bracket that are preserved under the group multiplication.

The structure of the letter is as follows. In the next section we introduce a novel Poisson-Lie group approach to the quadratic Hamiltonian structure of LV equations. In Section 3 we use the group multiplication map $\Delta^{(2)}$ in order to obtain new integrable perturbations of the 3D LV equations through the embedding method. In Section 4 the most generic deformation of the quadratic Poisson-Lie structure is found, and a new two-parametric integrable deformation of the LV equations containing polynomial and rational perturbation terms is explicitly obtained. A final section summarizing the main results and pointing out several open questions closes the paper.


\sect{Lotka-Volterra equations as Poisson-Lie dynamics}

Let us consider the following quadratic Poisson algebra ${\cal P}$~\cite{CDamianou, ChrisDamianou} that generalizes (\ref{pnut})
  \be
\begin{array}{lll}
\{X,Y\}=\alpha XY\qquad & \{X,Z\}= \beta X Z \qquad & \{Y,Z\}=\gamma Y Z.
\end{array}
\label{palg}
\ee
The Casimir function for ${\cal P}$ reads
\be
\mathcal{C}=X^{-\gamma} Y^{\beta} Z^{-\alpha}.
\ee
If we consider the Hamiltonian function -which also generalizes (\ref{hamab})-
\be
\mathcal{H}=a_{1}\,X+ a_{2}\,Y+a_{3}\,Z+b_{1}\,\log X+b_{2}\,\log Y +b_{3}\,\log Z
\label{LVh1}
\ee
we get the following LV equations as the integrable dynamical system given by $
\dot{F}=\{F, \mathcal{H}\} 
$:
\begin{eqnarray}
\dot{X}&=& X[\alpha a_{2}\, Y+\beta a_{3}\, Z+ (\alpha b_{2}+\beta b_{3})]\notag\\
\dot{Y}&=&Y[
-\alpha a_{1}\,X+\gamma a_{3}\, Z+ (\gamma b_{3}-\alpha b_{1})
]\label{lvgen}\\
\dot{Z}&=& Z[
-\beta a_{1}\, X-\gamma a_{2}\,Y -(\beta b_{1}+\gamma b_{2})
].
\notag
\end{eqnarray}

These are Lotka-Volterra equations that depend on 9 free parameters, and the system is always integrable since the Hamiltonian and the Casimir function are constants of the motion.
Note that if the three $a_i$ parameters are all of them different from zero, then they can be reabsorbed in the Hamiltonian and equations through a linear change of variables, and  the following skew-symmetric form of the LV system is obtained (see~\cite{CDamianou, ChrisDamianou})
\begin{eqnarray}
\dot{X}&=& X[\alpha \, Y+\beta \, Z+ (\alpha b_{2}+\beta b_{3})]\notag\\
\dot{Y}&=&Y[
-\alpha \,X+\gamma \, Z+ (\gamma b_{3}-\alpha b_{1})
]\\
\dot{Z}&=& Z[
-\beta \, X-\gamma \,Y -(\beta b_{1}+\gamma b_{2})
].
\notag
\end{eqnarray}
Moreover, this system can be also shown to be equivalent (through a further linear change of variables) to the more usual 6-parameter LV equations given by (\ref{abc}) (the so-called ABC-system). In fact, if $\alpha,\beta,\gamma\neq 0$ and we define
\be
\begin{array}{lll}
a_{1}=-\dfrac{1}{\alpha}, & a_{2}=-\dfrac{1}{\gamma}, & a_{3}=\dfrac{1}{\beta}
\end{array}
\ee
then from (\ref{lvgen}) we get the ABC system with
\be
\begin{array}{lll}
A=\dfrac{\gamma}{\beta}, & B=\dfrac{\beta}{\alpha}, & C=-\dfrac{\alpha}{\gamma}\\
\lambda=\alpha b_{2}+\beta b_{3},\quad & \mu=-\alpha b_{1}+\gamma b_{3},\quad  & \nu=-\beta b_{1}-\gamma b_{2}
\end{array}
\ee
which fulfils the conditions 
\be
A\, B\, C=-1, \qquad \nu=\mu B-\lambda A B
\ee
that characterize the bi-Hamiltonian case described in~\cite{Nutku}.

At this point it is important to stress that, as we shall see in Section 3, after deforming the Poisson bracket (\ref{palg}) the abovementioned linear changes of variables that relate the different presentations of LV systems will no longer work. Therefore in the rest of the paper we will consider as the LV system the 9-parameter fully generic case (\ref{lvgen}).

\subsect{The LV Poisson algebra as a Poisson-Lie group}

Let us consider  the Lie group $G$ generated by the 3D multiparametric  Lie algebra $g_{\alpha,\beta,\gamma}$ given by
\be
[e_0,e_1]=\frac{\gamma}{\beta}\,e_1 \qquad 
[e_0,e_2]=-\frac{\gamma}{\alpha}\,e_2  \qquad [e_1,e_2]=0.
\ee
A three-dimensional representation of this Lie algebra is given by:
\be
\rho(e_1)=\left( 
\begin{array}{ccc}
0 & 0 & 1\\
0 & 0 & 0\\
0 & 0 & 0
\end{array}
\right),  \qquad 
\rho(e_2)=\left( 
\begin{array}{ccc}
0 & 0 & 0\\
0 & 0 & 1\\
0 & 0 & 0
\end{array}
\right), \qquad
\rho(e_0)=\left( 
\begin{array}{ccc}
\frac{\gamma}{\beta} & 0 & 0\\
0 & -\frac{\gamma}{\alpha} & 0\\
0 & 0 & 0
\end{array}
\right).
\ee
In this representation, the three-dimensional form of a generic Lie group element $M$ reads:
\be
M=\exp \left(E_1 \rho(e_1) \right) \exp \left(E_2 \rho(e_2) \right) \exp \left(E_0 \rho(e_0) \right)=
\left( 
\begin{array}{ccc}
\exp(\tfrac{\gamma}{\beta}\,E_0) & 0 & E_1\\
0 & \exp(-\tfrac{\gamma}{\alpha}\,E_0) & E_2\\
0 & 0 & 1
\end{array}
\right).
\ee
By changing to the new group coordinates $X=\exp(E_0), Y=E_1, Z=E_2$, we obtain
\be
M=\left(
\begin{array}{ccc}
 X^{\frac{\gamma}{\beta}} & 0 & Y \\
 0 & X^{-\tfrac{\gamma}{\alpha}} & Z\\
 0 & 0 & 1
\end{array} 
\right)
\label{group}
\ee
and the matrix product of two different group elements $M_2$ and $M_1$ will be
\be
M_2\cdot M_1=\left(
\begin{array}{ccc}
 X_2^{\frac{\gamma}{\beta}} & 0 & Y_2 \\
 0 & X_2^{-\tfrac{\gamma}{\alpha}} & Z_2\\
 0 & 0 & 1
\end{array} 
\right)
\cdot
\left(
\begin{array}{ccc}
 X_1^{\frac{\gamma}{\beta}} & 0 & Y_1 \\
 0 & X_1^{-\tfrac{\gamma}{\alpha}} & Z_1\\
 0 & 0 & 1
\end{array} 
\right)
=\left(
\begin{array}{ccc}
 (X_2\,X_1)^{\frac{\gamma}{\beta}} & 0 & X_2^{\frac{\gamma}{\beta}}\,Y_1 + Y_2\\
 0 & (X_2\,X_1)^{-\tfrac{\gamma}{\alpha}} & X_2^{-\tfrac{\gamma}{\alpha}}\,Z_1 +Z_2 \\
 0 & 0 & 1
\end{array} 
\right). 
\label{producto}
\ee

By following the usual convention in quantum group theory (see~\cite{CP} for details), this group multiplication can be rewritten in more algebraic terms by  identifying the two sets of group coordinates
$\{X_1,Y_1,Z_1\}$ and $\{X_2,Y_2,Z_2\}$ with the `tensor product variables'
\begin{eqnarray}
& X_1= X \otimes 1, \qquad X_2= 1 \otimes X & \nonumber \\ 
& Y_1= Y \otimes 1, \qquad Y_2= 1 \otimes Y & \label{natcoord} \\
& Z_1= Z \otimes 1, \qquad Z_2= 1 \otimes Z & \nonumber
\end{eqnarray} 
that come from considering each set of group coordinates as a set of dynamical variables living on a different copy of the group manifold. Under this viewpoint, the product of two group elements reads
\be
M_2\cdot M_1=\left(
\begin{array}{ccc}
 X^{\frac{\gamma}{\beta}}\otimes X^{\frac{\gamma}{\beta}} & 0 & \,Y\otimes X^{\frac{\gamma}{\beta}}+ 1\otimes Y\\
 0 & X^{-\tfrac{\gamma}{\alpha}}\otimes X^{-\tfrac{\gamma}{\alpha}} & Z\otimes X^{-\tfrac{\gamma}{\alpha}}\, +1\otimes Z\\
 0 & 0 & 1\otimes 1
\end{array} 
\right). 
\label{coproducto}
\ee
This algebraic approach suggests that the matrix elements of the product (\ref{coproducto}) of two group elements can be interpreted as the images of a mapping $\Delta^{(2)}:M\rightarrow M\otimes M$, namely
\be
M_2\cdot M_1=\left(
\begin{array}{ccc}
 [\Delta^{(2)}(X)]^{\frac{\gamma}{\beta}} & 0 &  \Delta^{(2)}(Y) \\
 0 &  [\Delta^{(2)}(X)]^{-\tfrac{\gamma}{\alpha}} &  \Delta^{(2)}(Z)\\
 0 & 0 & \Delta^{(2)}(1)
\end{array} 
\right).
\label{coproducto2}
\ee
Such mapping is called the coproduct (or comultiplication map) and reads
\bea
&& \Delta^{(2)}(X)=X\otimes X=X_{2}X_{1}\notag\\
&& \Delta^{(2)}(Y)=Y\otimes X^{\frac{\gamma}{\beta}}+ 1\otimes Y=X_{2}^{\frac{\gamma}{\beta}}Y_{1}+Y_{2}\\
&& \Delta^{(2)}(Z)=Z\otimes X^{-\tfrac{\gamma}{\alpha}}\, +1\otimes Z=X_{2}^{-\frac{\gamma}{\alpha}}Z_{1}+Z_{2},
\notag
\eea
besides $\Delta^{(2)}(1)=1\otimes 1$. Obviously, this construction can be generalized to the product of $N$ group elements, and then we get the explict form of the $N$-th coproduct $\Delta^{(N)}$, that maps $M$ into the tensor product of $N$ copies of $M$.

Now, the key observation is the following invariance property: If we assume that the `abstract' commutative algebra of smooth functions on the group coordinates $\{X,Y,Z\}$ is endowed with the LV Poisson structure $\mathcal{P}$ given by (\ref{palg}), then it turns out that the coproduct $\Delta^{(2)}$ is a Poisson algebra homomorphism with respect to $\mathcal{P}$, {\em i.e.}, the following relations hold:
 \bea
&&
\{\Delta^{(2)}(X), \Delta^{(2)}(Y)\}=\alpha\, \Delta^{(2)}(X)\, \Delta^{(2)}(Y)\notag\\
 && \{\Delta^{(2)}(X),\Delta^{(2)}(Z)\}= \beta\,\Delta^{(2)}(X) \,\Delta^{(2)}(Z) \label{co2} \\
 && \{ \Delta^{(2)}(Y),\Delta^{(2)}(Z)\}=\gamma\, \Delta^{(2)}(Y)\, \Delta^{(2)}(Z),
 \notag
\eea
where in (\ref{co2}) $\{, \}$ denotes the natural Poisson structure on $\mathcal{P} \otimes \mathcal{P}$ given by
\be
\{ a \otimes b, c \otimes d\}=\{a,c \} \otimes bd +ac \otimes \{ b, d \}. 
\label{pois2}
\ee

The previous statement is proven by a straightforward computation and means that $(C^{\infty}(G),\mathcal{P})$ is a Poisson-Lie group (equivalently, that $(\mathcal{P},\Delta^{(2)})$ is a Poisson coalgebra~\cite{CP,BR}).
Therefore, from this perspective we can conclude that the LV equations (\ref{lvgen}) are just a specific example of Poisson-Lie dynamics defined on $(C^{\infty}(G),\mathcal{P})$ and generated by the Hamiltonian
\be
\mathcal{H}=a_{1}\,X+ a_{2}\,Y+a_{3}\,Z+b_{1}\,\log X+b_{2}\,\log Y +b_{3}\,\log Z.
\ee
Therefore, this result endows the known quadratic LV bracket (\ref{palg}) with a new underlying Poisson-Lie group structure, and the coproduct map $\Delta^{(2)}$ will be essential in order to get integrable deformations of the LV system.


\sect{Integrable deformations induced from the coproduct map}

In the sequel we show that  the Poisson-Lie nature of the bracket $\mathcal{P}$ provides a straightforward deformation approach to LV equations which is based on their embedding within higher dimensional integrable systems. These systems will be obtained by taking into account that Poisson algebras endowed with a coproduct map $\Delta^{(2)}$ give rise to a systematic way of constructing integrable systems with an arbitrary number of degrees of freedom   (for a detailed exposition of such coalgebra approach to integrability, including its relation with other approaches to integrability and many explicit examples, see~\cite{BR,RutwigProc, loop1, loop2}). 

In order to make this construction explicit, let us work out the $N=2$ case by considering as the new Hamiltonian $\mathcal{H}^{(2)}$ the coproduct $\Delta^{(2)}$ of the $N=1$ Hamiltonian (\ref{LVh1}), namely
\begin{eqnarray}
\!\!\!\!\!\!\!\!\!\!
&& \mathcal{H}^{(2)}:=\Delta^{(2)}(\mathcal{H})\notag\\
&& =
a_{1}\,\Delta^{(2)}(X)+ a_{2}\,\Delta^{(2)}(Y)+a_{3}\,\Delta^{(2)}(Z)+b_{1}\,\Delta^{(2)}(\log X)+b_{2}\,\Delta^{(2)}(\log Y) +b_{3}\,\Delta^{(2)}(\log Z)
\notag\\
&& = a_{1}\left(X_{2}X_{1}\right)+a_{2}\left(
X_{2}^{\frac{\gamma}{\beta}}Y_{1}+Y_{2}
\right)+a_{3}\left(X_{2}^{-\frac{\gamma}{\alpha}}Z_{1}+Z_{2}\right)\notag\\
&& \qquad +b_{1}\log\left( X_{2}X_{1}\right)+b_{2}\log \left(
X_{2}^{\frac{\gamma}{\beta}}Y_{1}+Y_{2}
\right)+b_{3}\log \left(
X_{2}^{-\frac{\gamma}{\alpha}}Z_{1}+Z_{2}
\right).
\label{ham2}
\end{eqnarray}
Now, by considering the natural definition (\ref{pois2}) of the Poisson bracket  on two copies of the group manifold,
we get the following Hamilton equations for this 6D dynamical system:
\small
\begin{eqnarray}
\dot{X}_{1}&=&X_{1}\left(\alpha\, a_{2}Y_{1}X_{2}^{\frac{\gamma}{\beta}} +\beta\, a_{3}Z_{1}X_{2}^{-\frac{\gamma}{\alpha}}\right)+X_{1}\left[
\alpha Y_{1}\left(\frac{b_{2}X_{2}^{\frac{\gamma}{\beta}}}{X_{2}^{\frac{\gamma}{\beta}}Y_{1}+Y_{2}}\right)+\beta Z_{1}\left(
\frac{b_{3}X_{2}^{-\frac{\gamma}{\alpha}}}{X_{2}^{-\frac{\gamma}{\alpha}}Z_{1}+Z_{2}}
\right)
\right]\notag\\
\dot{Y}_{1}&=&Y_{1}\left(
-\alpha a_{1}X_{1}X_{2}+\gamma a_{3}Z_{1}X_{2}^{-\frac{\gamma}{\alpha}}
\right)+\gamma Y_{1}\left[
\frac{b_{3}Z_{1}X_{2}^{-\frac{\gamma}{\alpha}}}{X_{2}^{-\frac{\gamma}{\alpha}}Z_{1}+Z_{2}}
\right]-\alpha\, b_{1} Y_{1}\notag\\
\dot{Z}_{1}&=&Z_{1}\left(
-\beta\, a_{1} X_{1}X_{2}-\gamma Y_{1}X_{2}^{\frac{\gamma}{\beta}}
\right)-\gamma\, Z_{1}\left[
\frac{b_{2}Y_{1}X_{2}^{\frac{\gamma}{\beta}}}{X_{2}^{\frac{\gamma}{\beta}}Y_{1}+Y_{2}}
\right]-\beta\, b_{1}Z_{1}\notag\\
\dot{X}_{2}&=& X_{2}(\alpha a_{2} Y_{2}+\beta a_{3} Z_{2})+X_{2}\left[\alpha Y_{2}\left(
\dfrac{b_{2}}
{
X_{2}^{\frac{\gamma}{\beta}}Y_{1}+Y_{2}
}
\right)+\beta Z_{2}\left(
\dfrac{b_{3}}{X_{2}^{-\frac{\gamma}{\alpha}}Z_{1}+Z_{2}}
\right)\right]
\notag\\
\dot{Y}_{2}&=&Y_{2}(-\alpha a_{1}X_{2}X_{1}+\gamma a_{3}Z_{2})+\gamma Y_{2}\left[
\dfrac{b_{3}\, Z_{2}}{X_{2}^{-\frac{\gamma}{\alpha}}Z_{1}+Z_{2}}
\right]\notag\\
&& -\alpha Y_{2}\left[
b_{1}+\dfrac{\gamma}{\beta}X_{2}^{\frac{\gamma}{\beta}}Y_{1}\left(
a_{2}+\dfrac{b_{2}}
{X_{2}^{\frac{\gamma}{\beta}}Y_{1}+Y_{2}}
\right)-\dfrac{\gamma}{\alpha}X_{2}^{-\frac{\gamma}{\alpha}}Z_{1}\left(
a_{3}+\dfrac{b_{3}}{X_{2}^{-\frac{\gamma}{\alpha}}Z_{1}+Z_{2}}
\right)
\right]\notag\\
\dot{Z}_{2}&=&Z_{2}(-\beta a_{1}X_{2}X_{1}-\gamma a_{2}Y_{2})-
\gamma Z_{2}\left[
\dfrac{b_{2}\, Y_{2}}
{
X_{2}^{\frac{\gamma}{\beta}}Y_{1}+Y_{2}
}
\right]\notag\\
&& -\beta Z_{2}\left[
b_{1}+\dfrac{\gamma}{\beta}X_{2}^{\frac{\gamma}{\beta}}Y_{1}\left(
a_{2}+\dfrac{b_{2}}
{X_{2}^{\frac{\gamma}{\beta}}Y_{1}+Y_{2}}
\right)-\dfrac{\gamma}{\alpha}X_{2}^{-\frac{\gamma}{\alpha}}Z_{1}\left(
a_{3}+\dfrac{b_{3}}{X_{2}^{-\frac{\gamma}{\alpha}}Z_{1}+Z_{2}}
\right)
\right]
\end{eqnarray}
\normalsize

This system is completely integrable since by making use of the Poisson coalgebra approach it is immediate to show~\cite{BR} that, besides the  Hamiltonian $\mathcal{H}^{(2)}$, there exist three more integrals of the motion in involution. They are the two Casimir functions for each subset of group variables together with the coproduct of the Casimir function, namely:
\bea
&& \mathcal{C}_{1}=X_{1}^{-\gamma}Y_{1}^{\beta}Z_{1}^{-\alpha}\qquad\qquad\qquad
\mathcal{C}_{2}=X_{2}^{-\gamma}Y_{2}^{\beta}Z_{2}^{-\alpha}
\\
&&\mathcal{C}^{(2)}:= \Delta^{(2)}(\mathcal{C})=\left(X_{1}X_{2}\right)^{-\gamma}\left(
X_{2}^{\frac{\gamma}{\beta}}Y_{1}+Y_{2}
\right)^{\beta}\left(
X_{2}^{-\frac{\gamma}{\alpha}}Z_{1}+Z_{2}
\right)^{-\alpha}.
\eea

It is straightforward to realize that the previous six equations are two `twisted' sets of integrable deformations of the 3D LV system. In fact, the equations for $\{ X_{2}, Y_{2}, Z_{2}\}$ are just an integrable deformation of the LV equations provided that the constraint $\dot{X}_{1}=0$ is imposed, and the perturbation terms contain both the $\{ X_{2}, Y_{2}, Z_{2}\}$ variables as well as the $\{Y_{1}, Z_{1}\}$ ones. Complementarily, the equations for $\{ X_{1}, Y_{1}, Z_{1}\}$ are another integrable deformation the of LV equations under the constraint $\dot{X}_{2}=0$. Obviously, this dynamical twisting comes indeed from the explicit form of the coproduct map $\Delta^{(2)}$ or, in geometric terms, from the fact that the group law that defines the coproduct is non-abelian.

Higher dimensional generalizations of this construction are straightforwardly obtained from the definition of the $N$-th coproduct map $\Delta^{(N)}$.
By following the same procedure as in the $N=2$ case -see (\ref{producto}) and (\ref{coproducto})-, the $N$-th coproduct map is obtained by computing the product of $N$ group matrices $M_N\cdot M_{N-1}\cdot\dots\cdot M_2\cdot M_1$ and reads
\be
\Delta^{(N)}(X)=\prod\limits_{i=1}^{N}X_{i}\qquad
\Delta^{(N)}(Y)=\sum\limits_{i=1}^{N}Y_{i}\prod\limits_{j=i+1}^{N}X_{j}^{\gamma/ \beta}
\qquad\Delta^{(N)}(Z)=\sum\limits_{i=1}^{N}Z_{i}\prod\limits_{j=i+1}^{N}X_{j}^{-\gamma/\alpha}
\label{Ncop}
\ee
which is again, by construction, a Poisson homomorphism for the  $N$-th tensor product of the LV bracket $\mathcal{P}$, as it can be explicitly checked.
Now, if we define the Hamiltonian $\mathcal{H}^{(N)}$ as the $N$-th coproduct of $\mathcal{H}$
\be
\mathcal{H}^{(N)}=a_{1}\,\Delta^{(N)}(X)+ a_{2}\,\Delta^{(N)}(Y)+a_{3}\,\Delta^{(N)}(Z)+b_{1}\,\Delta^{(N)}(\log X)+b_{2}\,\Delta^{(N)}(\log Y) +b_{3}\,\Delta^{(N)}(\log Z)
\label{hamn}
\ee
we get a set of 3N Hamilton equations for the variables $\{ X_{k}, Y_{k}, Z_{k}\}$, that we do not write explicitly here for the sake of brevity. As expected, this system generalizes  the $N=2$ pattern previuosly described: for each fixed subset of three variables $\{ X_{k}, Y_{k}, Z_{k}\}$ we recover an integrable deformation of the LV system provided that the $(N-1)$ constraints $\dot{X}_{j}=0$ for all $j \neq k$ are imposed. These constraints reflect again the `twisting' between the $N$ different sets of perturbed LV equations that arise as a consequence of the Poisson invariance of the system under the $N$-th coproduct map. We recall that the generic properties of the dynamical embedding of a $(N-1)$-th coproduct system into the corresponding $N$-th coproduct one were discussed in~\cite{cluster} in terms of the so-called `cluster variables'.

Finally, by working out in detail the general formalism of Poisson coalgebra integrability given in~\cite{BR}, it can be proven that the set of integrals of the motion in involution for this $3\,N$-dimensional system are the Hamiltonian $\mathcal{H}^{(N)}$ together with the $N+(N-1)$ functions given by the Casimirs on each group element and the $r$-th coproducts $(r=2,\dots,N)$ of the abstract Casimir $\mathcal{C}$. Explicitly,
\bea
&&\!\!\! \!\!\!   \mathcal{C}_{i}=X_{i}^{-\gamma}Y_{i}^{\beta}Z_{i}^{-\alpha}\qquad\qquad\qquad
i=1,\dots,N\label{casni}\\
&&\!\!\! \!\!\! \mathcal{C}^{(r)}:= \Delta^{(r)}(\mathcal{C})=
[\Delta^{(r)}(X)]^{-\gamma}\,
[\Delta^{(r)}(Y)]^{\beta}\,
[\Delta^{(r)}(z)]^{-\alpha}\notag\\
&&\qquad\qquad
=\left[ \prod\limits_{i=1}^{r}X_{i} \right]^{-\gamma}\,
\left[ \sum\limits_{i=1}^{r}Y_{i}\prod\limits_{j=i+1}^{r}X_{j}^{\gamma/ \beta} \right]^{\beta}\,
\left[\sum\limits_{i=1}^{r}Z_{i}\prod\limits_{j=i+1}^{r}X_{j}^{-\gamma/\alpha} \right]^{-\alpha}
 r=2,\dots,N
\label{cocasni}
\eea
where we have used the superscript $(r)$ to denote the comultiplication map between $M$ and the tensor product of $r$ copies of $M$.

\sect{Integrable perturbations from a deformed Poisson-Lie group}

This group-theoretical interpretation of the LV equations (\ref{lvgen}) raises also the natural question concerning the existence of deformations of the Poisson structure $\mathcal{P}$  (\ref{palg}) that could be also invariant under the coproduct map $\Delta^{(2)}$. Equivalently, this is just the problem of finding other PL structures on the group $G$, whose existence would provide new integrable deformations of the LV equations.

This question can be answered in the affirmative, and the following result can be obtained through direct computation: 

\noindent {\bf Proposition}. The most generic quadratic Poisson structure in $\{X,X^{\frac{\gamma}{\beta}},X^{-\frac{\gamma}{\alpha}},Y,Z,1\}$ and for which the comultiplication $\Delta^{(2)}$ is a Poisson map is given by the brackets
 \begin{eqnarray}
\{X,Y\}&=&\alpha X Y +\delta X(1-X^{\frac{\gamma}{\beta}})\notag\\
 \{X,Z\}&=&\beta X Z +\epsilon X(1-X^{-\frac{\gamma}{\alpha}})\label{poisde}\\
\{Y,Z\}&=&\gamma Y Z +\dfrac{\gamma \epsilon}{\beta}Y+\dfrac{\gamma \delta}{\alpha}Z+\dfrac{\gamma \delta \epsilon}{\beta \alpha}\left(
1 -X^{\frac{\gamma}{\beta}-\frac{\gamma}{\alpha}}
\right).\notag
\end{eqnarray}
We shall call this Poisson bivector as ${\cal P}_{\delta,\epsilon}$. Moreover, the Casimir function for ${\cal P}_{\delta,\epsilon}$ is found to be
\be
\mathcal{C}_{\delta,\epsilon}=
\left[
\delta (1-X^{\frac{\gamma}{\beta}})+\alpha Y
\right]^{-\frac{\beta}{\alpha}}\left[
\epsilon (X^{\frac{\gamma}{\alpha}}-1)+ \beta Z X^{\frac{\gamma}{\alpha}}
\right].
\ee
\medskip
Consequently, we can say that $({\cal P}_{\delta,\epsilon},\Delta^{(2)})$ is a multiparametric PL structure on $G$  whose limit $\delta,\epsilon\to 0$ leads to the LV Poisson structure $\mathcal{P}$.

With this result in mind, an integrable $(\delta,\epsilon)$-deformation of the LV equations is straightforwardly obtained by considering again the same Hamiltonian (\ref{LVh1})
and the new Poisson bracket ${\cal P}_{\delta,\epsilon}$. In this way we obtain the perturbed LV equations
\smallskip
\begin{eqnarray}
\dot{X}&=&X \left[
\alpha a_{2} Y+\beta a_{3} Z+ (\alpha b_{2}+\beta b_{3})\right]\notag\\
&& \qquad +\delta X \left(
1-X^{\frac{\gamma}{\beta}}
\right)\left(a_{2}+\dfrac{b_{2}}{Y}\right)+\epsilon X  \left(
1-X^{-\frac{\gamma}{\alpha}}
\right)\left(a_{3}+\dfrac{b_{3}}{Z}\right)
\notag\\
\dot{Y}&=&Y\left[
-\alpha a_{1}X+\gamma a_{3}Z+(-\alpha b_{1}+\gamma b_{3})\right]+\delta \left[(X^{\frac{\gamma}{\beta}}-1)(a_{1}X+b_{1})+\dfrac{\gamma}{\alpha}\left(a_{3}Z+b_{3}\right)\right]
\notag\\
&&+\dfrac{\epsilon \gamma}{\beta} \left[
Y\left(a_{3}+\dfrac{b_{3}}{Z}\right)+\dfrac{\delta}{\alpha}\left(
a_{3}+\dfrac{b_{3}}{Z}
\right)\left(1-X^{\frac{\gamma}{\beta}-\frac{\gamma}{\alpha}}\right)
\right]\label{LVed}\\
\dot{Z}&=&Z\left[
-\beta a_{1}X-\gamma a_{2}Y+(-\gamma b_{2}-\beta b_{1})\right]-\dfrac{\delta \gamma}{\alpha}Z\left[
a_{2}+\dfrac{b_{2}}{Y}
\right]\notag\\
&& +\epsilon \left[
(X^{-\frac{\gamma}{\alpha}}-1)(a_{1}X+b_{1})-\dfrac{\gamma}{\beta}\left(
a_{2}\,Y+b_{2}
\right)
\right]
+ \dfrac{\delta \epsilon \gamma}{\alpha \beta }\left[
(X^{\frac{\gamma}{\beta}-\frac{\gamma}{\alpha}}-1)\left(
a_{2}+\dfrac{b_{2}}{Y}
\right)
\right]\notag
\end{eqnarray}
\normalsize 
whose integrability is preserved despite the presence of both polynomial and rational perturbation terms, since both $\mathcal{H}$ and the deformed Casimir $\mathcal{C}_{\delta,\epsilon}$ are integrals of the motion in involution for the system. 

Obviously, the $(\delta,\epsilon)$-integrable deformation of the $N$ sets of twisted LV equations given in Section 3 can be  obtained by using ${\cal P}_{\delta,\epsilon}$ (instead of $\mathcal{P}$) and by making use of the same $N$-th coproduct map (\ref{Ncop}) and Hamiltonian (\ref{hamn}) since ${\cal P}_{\delta,\epsilon}$ is invariant under the same group multiplication. Again we do not present the explicit form of the set of 3N equations here, but we would like to stress that in this deformed case the following (2N-1) integrals of the motion for the system can be explicitly obtained from the coalgebra integrability approach:
\be
\mathcal{C}_{i}=
\left[
\delta \left(1-X_{i}^{\frac{\gamma}{\beta}}\right)+\alpha Y_{i}
\right]^{-\frac{\beta}{\alpha}}\left[
\epsilon \left(X_{i}^{\frac{\gamma}{\alpha}}-1\right)+\beta Z_{i} X_{i}^{\frac{\gamma}{\alpha}}
\right]
\,\,\,\,\,\,i=1,2, \ldots, N
\ee
\be
\Delta^{(r)}(C)=
\left[
\delta \left(1-\Delta^{(r)}(X)^{\frac{\gamma}{\beta}}\right)+\alpha \Delta^{(r)}(Y)
\right]^{-\frac{\beta}{\alpha}}\left[
\epsilon \left(\Delta^{(r)}(X)^{\frac{\gamma}{\alpha}}-1\right)+\beta \Delta^{(r)}(Z) \Delta^{(r)}(X)^{\frac{\gamma}{\alpha}}
\right]
\,\,\,r=2,\ldots, N.
\ee
Indeed, these are the appropriate deformations of the coalgebra integrals (\ref{casni}) and (\ref{cocasni}).


\section*{Concluding remarks}

Summarizing, we have presented a new Poisson-Lie interpretation of the quadratic Poisson structure that underlies the integrability of the 3D LV system and we have shown that this additional group invariance property is useful for getting new integrable deformations of the LV system through two different procedures. The first one makes use of the invariance of the Poisson bracket under the product of $N$ group elements, and gives rise to a nested set of $N$ deformed LV systems. The second one provides a two-parametric familly of integrable deformations of the 3D LV equations by finding the most generic deformation of the quadratic Poisson bracket that is compatible with the group multiplication. Therefore, this quantum-group algebraic machinery (we recall that Poisson-Lie groups are just the `semiclassical' limit of quantum groups~\cite{CP}) turns out to be again useful in a classical integrability context.

Concerning applications, it would be certainly intreresting to find specific LV models in which  the $(\delta,\epsilon)$-perturbation terms appearing (\ref{LVed}) could be dynamically meaningful, and in this context the presence of rational perturbation terms could give rise to dynamical regimes with saturation rates. Also, the generalization of the Poisson-Lie group approach here presented to the case of 4D (and higher dimensional) LV systems is under investigation.


\section*{Acknowledgements}

This work was partially supported by the Spanish MICINN   under grant   MTM2010-18556
and by INFN--MICINN.    




\begin{thebibliography}{99}

\bibitem{Lotka} A.J. Lotka, {\it Elements of Mathematical Biology}, (Dover: New York) (1956).

\bibitem{Volterra} V. Volterra, {\it Lecons sur la Th\'eorie Math\'ematique de la Lutte pour la Vie}, (Gauthier Villars: Paris) (1931).


\bibitem{Plank} M. Plank, J. Math. Phys. {\bf 36}, 3520 (1995).

\bibitem{BF} B. Hern\'andez-Bermejo, V. Fair\'en, Mathematical Biosciences {\bf 140}, 1 (1997).


\bibitem{Bountis} T.C. Bountis, M. Bier, J. Hijmans, Phys. Lett. A {\bf 97}, 11 (1983).

\bibitem{CFG} L. Cair\'o, M.R. Feix, J. Goedert, Phys. Lett. A {\bf 140}, 421 (1989).

\bibitem{Moulin} B. Grammaticos, J. Moulin-Ollagnier, A. Ramani, J.M. Strelcyn, S. Wojciechowski, Physica A {\bf 163}, 683 (1990).

\bibitem{CF} L. Cair\'o, M.R. Feix, J. Goedert, J. Math. Phys. {\bf 33}, 2440 (1992).

\bibitem{CLl} L. Cair\'o, J. Llibre, J. Phys. A: Math. Gen. {\bf 33}, 2395 (2000).

\bibitem{CDamianou} K. Constandinides, P.A. Damianou, Reg. Chaotic Dyn. {\bf 16}, 311 (2011).

\bibitem{ChrisDamianou} Y.T. Christodoulides, P.A. Damianou, J. Nonlin. Math. Phys. {\bf 16}, 339 (2009).


\bibitem{Nutku} Y. Nutku, Phys. Lett. A {\bf 145}, 27 (1990).

\bibitem{Damianou} P.A. Damianou, Phys. Lett. A {\bf 155}, 126 (1991).

\bibitem{GNutku} H. G\"umral, Y. Nutku, J. Math. Phys. {\bf 34}, 5691 (1993).

\bibitem{Plank2} M. Plank, Nonlinearity {\bf 9}, 887 (1996).

\bibitem{Kerner} E.H. Kerner, J. Math. Phys. {\bf 38}, 1218 (1997).

\bibitem{BF2} B. Hern\'andez-Bermejo, V. Fair\'en, J. Math. Phys. {\bf 39} 6162 (1998). 

\bibitem{Puyun} Puyun Gao, Phys. Lett. A {\bf 273}, 85 (2000).



\bibitem{CP} V. Chari, A. Pressley, {\it A Guide to Quantum Groups}, (Cambridge University Press: Cambridge) (1994).

\bibitem{BR} A. Ballesteros, O. Ragnisco, J. Phys. A: Math. Gen. {\bf 31}, 3791 (1998).

\bibitem{RutwigProc}
A. Ballesteros, A. Blasco, F.J. Herranz,  F. Musso,  O. Ragnisco, J. Phys: Conf. Ser. \textbf{175}, 012004
(2009).

\bibitem{loop1} F. Musso, J. Phys. A: Math. Theor. {\bf 43}, 434026 (2010).

\bibitem{loop2} F. Musso, J. Phys. A: Math. Theor. {\bf 43}, 455207 (2010).

\bibitem{cluster} A. Ballesteros, O. Ragnisco, J. Phys. A: Math. Gen. {\bf 36}, 10505 (2003).



\end{thebibliography}
\end{document}